\documentclass[final,1p,times]{elsarticle}

\usepackage{pifont}
\usepackage{natbib}
\usepackage{geometry}
\usepackage{graphicx}
\usepackage{txfonts}
\usepackage{subfigure}
\usepackage{wrapfig}
\usepackage{color}
\usepackage{booktabs}
\usepackage{caption}
\usepackage{amsmath}
\usepackage{amssymb}
\usepackage{url}
\usepackage{colortbl}
\usepackage{multirow}
\usepackage{fnbreak}
\usepackage{helvet}
\usepackage[font=bf]{caption}

\graphicspath{{figs/}}

\definecolor{Gray}{gray}{0.7}

\begin{document}

\begin{frontmatter}

\begin{center}
\small \textbf{9\textsuperscript{th} IAA Planetary Defense Conference -- PDC 2025} \\ \textbf{5--9 May 2025, Stellenbosch, Cape Town, South Africa} \\ \vspace{0.1in} \textbf{IAA-PDC25-04-164} \\
\end{center}

\title{Small Near-Earth Objects in the Taurid Resonant Swarm}  


\author[umd,bu]{Quanzhi Ye\corref{cor1}}
\ead{qye@umd.edu}
\author[pool,cmu]{Jasmine Li}
\author[uwo1,uwo2]{Denis Vida}
\author[uwo1,uwo2]{David L. Clark}
\author[uw]{Eric C. Bellm}

\cortext[cor1]{Corresponding author}

\address[umd]{Department of Astronomy, University of Maryland, College Park, MD 20742, USA}
\address[bu]{Center for Space Physics, Boston University, 725 Commonwealth Ave, Boston, MA 02215, USA}
\address[pool]{Poolesville High School, Poolesville, MD 20837, USA}
\address[cmu]{Department of Electrical and Computer Engineering, Carnegie Mellon University, Pittsburgh, PA 15289, USA}
\address[uwo1]{Department of Physics and Astronomy, University of Western Ontario, London, Ontario, N6A 3K7, Canada}
\address[uwo2]{Institute for Earth and Space Exploration, University of Western Ontario, London, Ontario N6A 5B8, Canada}
\address[uw]{DIRAC Institute, Department of Astronomy, University of Washington, 3910 15th Avenue NE, Seattle, WA 98195, USA}

\begin{abstract}
The Taurid Resonant Swarm (TRS) within the Taurid Complex hosts dynamically-concentrated debris in a 7:2 mean-motion resonance with Jupiter. Fireball observations have confirmed that the TRS is rich in sub-meter-sized particles, but whether this enhancement extends to larger, asteroid-sized objects remains unclear. Here we reanalyze the data obtained by a Zwicky Transient Facility (ZTF) campaign during the 2022 TRS encounter, and find that the TRS may host up to $\sim10^2$ Tunguska-sized objects and up to $\sim10^3$ Chelyabinsk-sized objects, the latter of which agrees the estimate derived from bolide records. This translates to an impact frequency of less than once every 4 million years. However, we caution that these numbers are based on the unverified assumption that the orbital distribution of the TRS asteroids follows that of fireball-sized meteoroids. Future wide-field facilities, such as the Vera C. Rubin Observatory, could take advantage of TRS's close approaches in the 2020--30s and validate the constraints of the asteroid-sized objects in the TRS.
\end{abstract}

\begin{keyword}
near-Earth objects \sep Taurid Resonant Swarm \sep impact risk
\end{keyword}

\end{frontmatter}

%
%

\section{Introduction}

The Taurid Resonant Swarm (TRS) is a dynamically coherent substructure within the broader Taurid Complex, a stream of debris associated with comet 2P/Encke and possibly a number of asteroids. Materials in the TRS are dynamically trapped in a 7:2 mean-motion resonance with Jupiter \cite{1993QJRAS..34..481A,1994VA.....38....1A,2019MNRAS.487L..35C}, resulting in episodic enhancements in the density of meteoroids and possibly larger near-Earth objects (NEOs) along specific orbital corridors. Earth periodically encounters the TRS, during which enhanced meteor activities have been observed \cite{1998MNRAS.297...23A,2017MNRAS.469.2077O}. Increased close encounters with larger asteroidal counterparts in TRS have also been predicted \cite{1994VA.....38....1A,2019MNRAS.487L..35C}, although unambiguous detections remain elusive.

Recent observational campaigns using wide-field telescopes, such as the 2022 Zwicky Transient Facility (ZTF) and Canada-France-Hawaii Telescope (CFHT) campaigns \cite{2025PSJ.....6...94L, 2025PSJ.....6..148W}, have sought to constrain the population of macroscopic TRS objects. The ZTF campaign placed an upper limit of no more than 9--14 objects brighter than an absolute magnitude of $H=24$ in TRS (equivalent to a diameter of $\gtrsim100$~m); the CFHT campaign set a $2\sigma$ upper limit of fewer than $(3$--$30)\times10^3$ objects down to $H=25.6\pm0.3$ (equivalent to a diameter of $\sim50$~m), suggesting a reduced impact hazard of hectometer-scale TRS bodies relative to earlier theoretical speculations. However, as shown by the Tunguska and Chelyabinsk events, smaller, decameter-scale NEOs can still cause regional devastation, yet they are small enough to evade current NEO surveys. Thus, it is important to quantify or at least constrain the abundance of these smaller NEOs in the TRS.

\section{Small NEOs in the TRS and Their Impact Risk}

The 2022 ZTF campaign searched an area of $\sim1600~\mathrm{deg^2}$ with an effective limiting magnitude of $V\sim20$, covering $99\%$ of $H=24$ TRS objects visible at the time of the campaign. The data were searched for moving objects using two approaches, targeting both trailed fast-moving and slower, non-trailed objects, with no candidates found in either approach.

To understand the population of small NEOs in the TRS, we reanalyze the results obtained in the 2022 ZTF campaign. We modify the $H$ of the synthetic population used to guide the ZTF campaign to cover TRS objects with $H=22$ to $27$, corresponding to diameters between $\sim250$~m to $10$~m, assuming 2P/Encke's albedo of $0.046$ \cite{2002EM&P...89..117C}. We then use the nondetection result as well as the detection efficiency established in \cite{2025PSJ.....6...94L} to calculate the upper limit of the number of objects in each $H$ bin.

\begin{figure*}[t!]
    \begin{center}
        \includegraphics[width=0.85\textwidth]{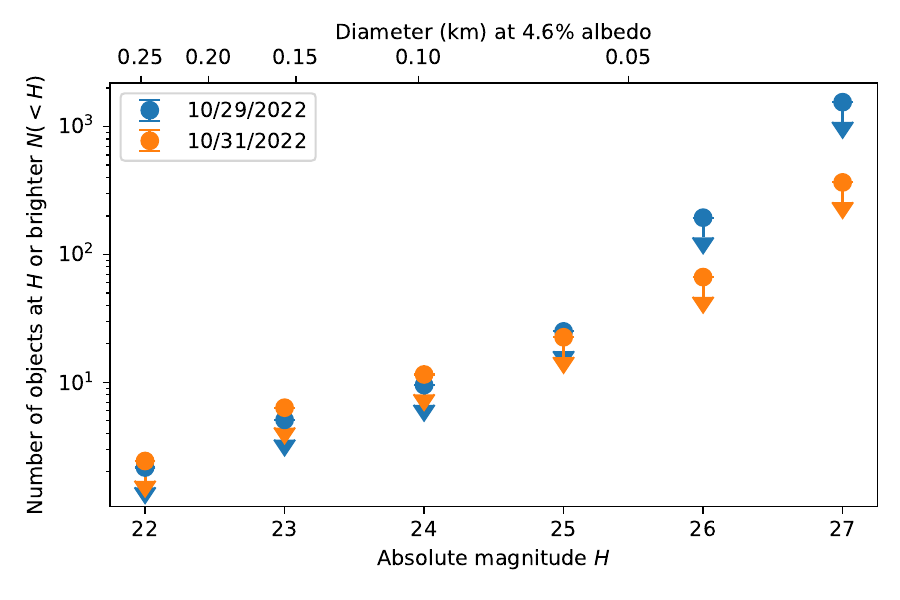}
        \caption{\small{Upper limits of TRS objects at an $H$ range of 22--27 based on ZTF campaign data obtained on the two campaign dates, 2022 October 29 and 31.}}
        \label{fig:number_h}
    \end{center}
\end{figure*}

Figure~\ref{fig:number_h} shows the upper limits of the number $N$ of TRS objects at $H$ range of 22--27. The distribution reaches a minimum at $N=1$ near $H\sim20-21$, which is in line with the largest NEO that is likely to be directly associated with the TRS, 2005 TF$_{50}$ ($H=20.3$) \cite{2021MNRAS.507.2568E}. At the smaller end, the 2022 October 31 data appears to provide a better constraint compared to the 2022 October 29 data, likely because Earth was closer to the center of the TRS on October 31. We determined that the TRS may host up to $\sim10^2$ Tunguska-sized objects (with diameters of $\sim50$~m) and up to $\sim10^3$ Chelyabinsk-sized objects (with diameters of $\sim20$~m). This is in line with the limit set by the 2022 CFHT campaign, which suggested no more than $(3$--$30)\times10^3$ objects down to Tunguska sizes \cite{2025PSJ.....6..148W}.

As a quick comparison, we search the Center for Near Earth Object Studies Fireball Database, a compilation of satellite-detected bolides from decameter-sized objects dating back to 1988\footnote{\url{https://cneos.jpl.nasa.gov/fireballs/}, also see \citet{2002Natur.420..294B} and \cite{2025Icar..42916444C}.}, for impactors possibly originating from the TRS. We look for bolides in the years of TRS encounters ($T_\mathrm{obs}=17$ years in total) with a loose constraint to budget for possible measurement errors: radiant within $30^\circ$ of the Taurids radiant, a speed within $30\%$ of the geocentric speed of the Taurids ($v_\mathrm{g}\approx30~\mathrm{km~s^{-1}}$), and arrival time within a month of the TRS crossing date, and do not find any candidates. The number of objects can then be estimated by

\begin{equation}
    N=\frac{L r_\mathrm{stream}^2}{R_\oplus^2 v_\mathrm{g}} \frac{N_\mathrm{obs}}{T_\mathrm{obs}}
\end{equation}

\noindent where $L\sim0.2$~au is the arc length of the Taurid orbit of a $\pm1$~month window, $r_\mathrm{stream}=0.01$~au is the (assumed) width of the TRS, $R_\oplus=6\times10^6$~m is the radius of the Earth, and $N_\mathrm{obs}=1$ is the (upper limit) number of TRS impactors. Subsituting these numbers into the equation, we obtain $N\sim10^2$, which is in line with the number we derived above on an order-of-magnitude level. This translates to a mass upper limit of $M\sim10^6$~kg for decameter-sized objects with a bulk density of $2000~\mathrm{kg~3^{-3}}$. We note that variables such as the observing window (and subsequently $L$) and $r_\mathrm{stream}$ are loosely constrained; for example, we get $N\sim10^4$ and $M\sim10^8$~kg if we use $r_\mathrm{stream}=0.1$~au, though this is still within the order of magnitude of the constraint set by the CFHT search.

The impact flux of the simulated TRS population can be estimated by

\begin{equation}
    F = n \sigma v_\mathrm{g}
\end{equation}

\noindent where $n$ is the number density of the TRS objects, $\sigma = \pi R_\oplus^2 \left( 1 + \frac{2GM_\oplus}{R_\oplus v_\mathrm{g}^2} \right)$ is the collision cross-section (with $G$, $R_\oplus$, $M_\oplus$ as the gravitational constant as well as the radius and mass of Earth). The number $n$ is derived from the simulated TRS population described in \cite{2025PSJ.....6...94L} by calculating the fraction of time each simulated particle spends within a sphere centered at 1~au (Earth's orbit) and dividing by the volume of the sphere, i.e.

\begin{equation}
    n = \frac{N \cdot \left\langle f \right\rangle}{4\pi (1~\mathrm{au})^2 \cdot \Delta r}
\end{equation}

\noindent where $N$ is the total number of TRS objects (upper limit) and $\left\langle f \right\rangle$ is the mean time fraction in the sphere of radius $\Delta r=10^{-4}$~au at 1~au. The value of $\left\langle f \right\rangle$ is found by taking the average of the sphere-crossing time of all simulated particles. This value of $n$ represents the number density of TRS objects near Earth's orbit.

Using $N\lesssim10^3$ from the upper limit of Chelyabinsk-sized objects in the TRS, we obtain $F\lesssim2\times10^{-7}~\mathrm{yr^{-1}}$, or less than one event every 4 million years. This is two orders of magnitude longer than the dynamical age of TRS, implying that an impact from a TRS object of Chelyabinsk-scale or larger is unlikely.

Although these numbers might seemingly ease the concern that TRS could produce many Earth impactors, we caution that they were derived based on the assumption that the orbital distribution of hypothetical TRS asteroids follows that of the observed TRS fireballs and meteors, which was inherited from the simulation used to guide the search. It is well known that larger bodies such as asteroids experience noticeably different dynamical processes compared to sub-meter-sized objects such as meteoroids. It is not clear at this point if this difference could lead to substantially different orbital outcomes that can affect the outcome of our calculation. Thus, further investigations into this matter as well as additional observational campaigns of the TRS asteroids are still desired.

\section{Future TRS Campaigns}

Earth will pass near the TRS again in 2025, 2026 and 2029, followed by two highly favorable, near-centric encounters in 2032 and 2036 \cite{1993QJRAS..34..481A}, which coincide in time with the commissioning of several large wide-field surveys, such as the Vera C. Rubin Observatory's Legacy Survey of Space and Time (LSST), offering the opportunity of conducting deeper searches along the TRS corridor.

The search for TRS objects differs from generic NEO searches, since the orbital space to search is much more defined and constrained. To compare the TRS search capability between different facilities, we devise a Figure of Merit (FoM) appropriate for streaked objects, originally derived by \cite{2017PASP..129c4402W}:

\begin{equation}
    \mathrm{FoM} \propto \left( \frac{\Omega_\mathrm{eff} 10^{0.6 m_\mathrm{lim}}}{t_\mathrm{obs}} \right) \left( \frac{\theta_\mathrm{PSF}}{\theta_\mathrm{streak}} \right)
\end{equation}

\noindent where $\Omega_\mathrm{eff}$ is the effective field of view to be accounted for the spatial size of the search corridor, $m_\mathrm{lim}$ is the effective limiting magnitude, $t_\mathrm{obs}$ is the total observation time including overheads, $\theta_\mathrm{PSF}$ is the pixel size of the point-source function (PSF), and $\theta_\mathrm{streak}$ is the angular length of a streak at the $m_\mathrm{lim}$ for a PSF source. The value of $\theta_\mathrm{streak}$ varies depending on the size, distance, and therefore detectability, of the targeted population; for ease of discussion, we use the median value of $6''$/min from the simulated TRS population for our calculation.

\begin{figure*}[t!]
    \begin{center}
        \includegraphics[width=\textwidth]{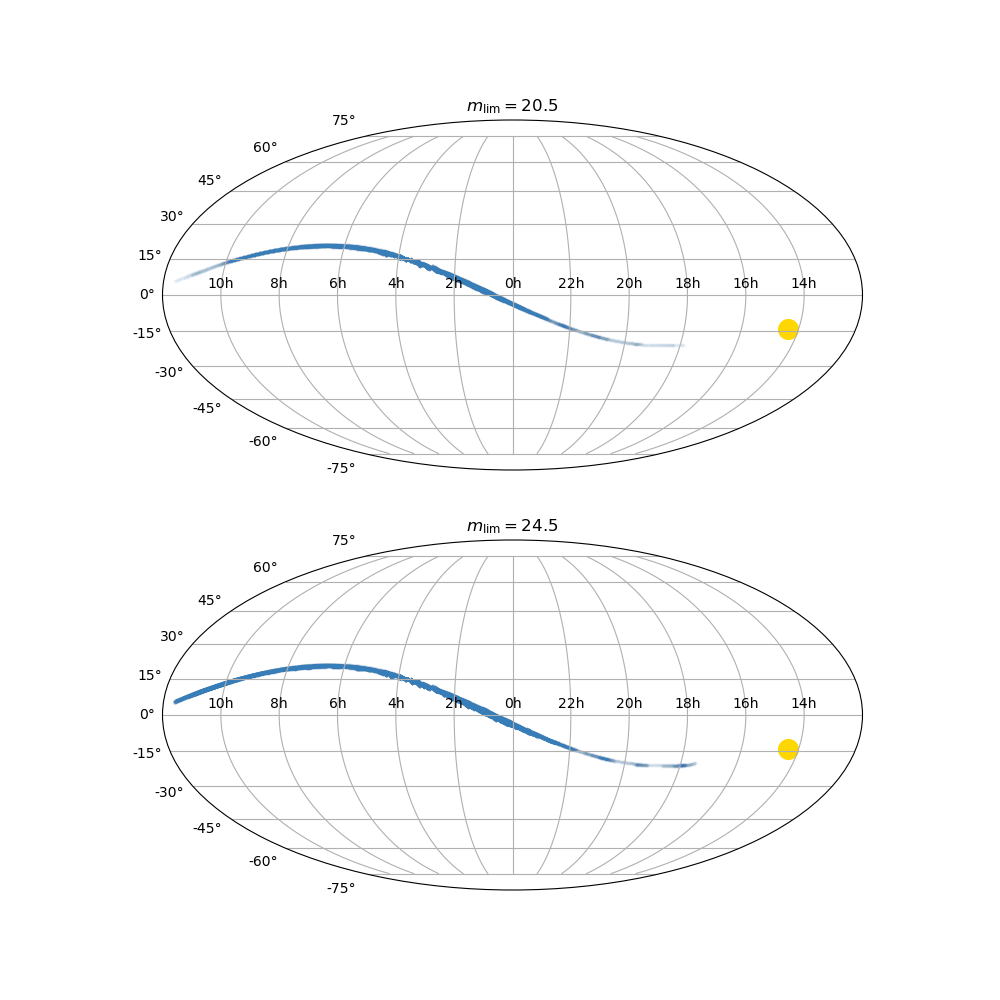}
        \caption{\small{On-sky footprints of simulated TRS particles on 2022 October 31 at ZTF ($m_\mathrm{lim}\sim20.5$) and LSST depth ($m_\mathrm{lim}\sim24.5$).}}
        \label{fig:footprint}
    \end{center}
\end{figure*}

The model-guided search corridor of the TRS asteroids during the close approach is a narrow arc of $\sim2^\circ$ in width circling almost the entire sky (Figure~\ref{fig:footprint}). Therefore, we have

\[
\Omega_\mathrm{eff}=
\begin{cases}
\displaystyle \Omega, &  (\Omega \leq 4~\mathrm{deg^2})\\
\displaystyle 2\sqrt{\Omega}, &  (\Omega > 4~\mathrm{deg^2})
\end{cases}
\]

\noindent where $\Omega$ is the original field of view of the telescope.

We calculate the FoM for a handful of wide-field telescopes, tabulated in Table~\ref{tab:trs_fom}. The advantage of LSST is significant: by this measure, LSST will be 28 times more efficient than ZTF for a TRS campaign. Telescopes with field-of-view significantly larger than $2^\circ$, such as ZTF, are actually at a disadvantage, since the TRS search corridor is only $\sim2^\circ$ in width.

\begin{table}[h]
\centering
\begin{tabular}{lcccccc}
\hline
Facility      & $m_{\rm lim}$ & $t_{\rm obs}$ (s) & $\Omega_{\rm eff}$ (deg$^2$) 
               & $\displaystyle\frac{\theta_{\rm PSF}}{\theta_{\rm streak}}$ 
               & $\mathrm{FoM}$ (rel.\ ZTF)\\
\hline
ZTF            & 20.6 & $(30+10)\times3$  & $2\sqrt{47.0}\approx13.7$& $0.33$  & 1  \\
LSST           & 24.5 & $(30+2)\times3$  & $2\sqrt{9.6}\approx6.2$   & $0.13$  & 28  \\
Subaru/HSC     & 24.4 & $(30+40)\times3$ & $1.8$                    & $0.06$ & 2.4   \\
DECam          & 23.3 & $(30+20)\times3$  & $3.0$                    & $0.06$   & 1.9  \\
CFHT/MegaCam   & 23.6 & $(30+12)\times3$ & $1.0$                    & $0.06$ & 0.8 \\
Pan-STARRS     & 22.7 & $(30+10)\times3$  & $2\sqrt{7.0}\approx5.3$  & $0.09$  & 1.8  \\
\hline
\end{tabular}
\caption{Comparison of the FoM of several wide-field telescopes. Listed are single‐visit effective limiting magnitude $m_{\rm lim}$, overhead-included observation time $t_{\rm obs}$ for 3 visits, effective field of view $\Omega_{\rm eff}$, and the streak‐reduction factor $\theta_{\rm PSF}/\theta_{\rm streak}$, with FoM normalized to $\mathrm{FoM_{ZTF}} = 1$. We assume the exposure time of single visit to be 30-s for all telescopes.}
\label{tab:trs_fom}
\end{table}

\begin{figure*}[t!]
    \begin{center}
        \includegraphics[width=0.65\textwidth]{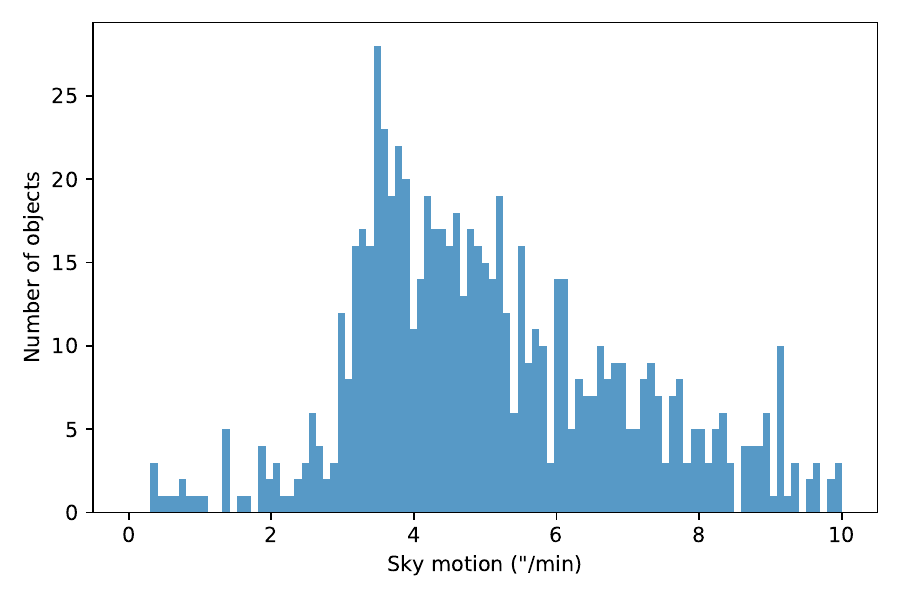}
        \caption{\small{On-sky motions of TRS objects on 2022 October 31.}}
        \label{fig:skymotion}
    \end{center}
\end{figure*}

One challenge for future TRS campaigns is the high sky motion of TRS objects during their close encounter to Earth. Taking the 2022 encounter for example, most TRS objects are moving at several arcsec/min or faster (Figure~\ref{fig:skymotion}), fast enough to streak on typical survey exposures. Facilities with larger pixel size, such as ZTF and LSST to a lesser extent, are less susceptible to the ``trailing loss'' caused by smeared objects. Detection algorithms specialized in identifying streaked objects will be critical in finding TRS objects in survey images.

\section{Conclusion}

We reanalyzed the 2022 ZTF campaign for TRS asteroids, and found that the TRS may host up to $\sim10^2$ Tunguska-sized objects and up to $\sim10^3$ Chelyabinsk-sized objects. The latter number is in line with estimate derived from bolide record on an order-of-magnitude level. The larger end of the size spectrum (near $H\sim20-21$) in our constraint appears to agree with the observed population. Our constraint translates into an impact frequency of Chelyabinsk-sized TRS objects of one event every 4 million years. These numbers should be taken with caution since they were derived under the assumption that TRS asteroids follow the same orbital distribution of smaller, fireball-sized TRS particles; whether this is true requires further investigation. Future multi-meter wide-field telescopes, such as the Simonyi Survey Telescope at the Vera C. Rubin Observatory, could enable a more sensitive search for TRS asteroids.

\section*{Acknowledgments}

This work is supported by NASA program 80NSSC22K0772. D.V. was supported in part by NASA Cooperative Agreement 80NSSC21M0073 and by the Natural Sciences and Engineering Research Council of Canada.

\bibliographystyle{model1-num-names}   
\bibliography{references}             

\vspace{0.25in}

%
%

\end{document}